\documentclass[dvips]{ws-procs9x6}
\usepackage{Ratcliffe}
\hypersetup{%
  pdfauthor={Philip G. Ratcliffe <philip.ratcliffe@uninsubria.it>},
  pdftitle={Transversity and Drell-Yan K-Factors},
  pdfsubject={Presentation at the XVI Int. Spin Symp., Trieste 2004},
  pdfkeywords={transversity, transverse spin, parton model},
}
\title{Transversity and Drell--Yan $K$-Factors}
\author{Philip G. Ratcliffe}
\address{%
    Dip.to di Fisica e Matematica,
    Universit\`{a} degli Studi dell'Insubria \\
  and \\
    Istituto Nazionale di Fisica Nucleare,
    Sezione di Milano
}
\begin{document}
\begin{fmffile}{Ratcliffe-fmf}
\maketitle
\renewcommand{\thefootnote}{\fnsymbol{footnote}}
\abstracts{%
  The Drell--Yan $K$-factors for transversely polarised hadrons are examined.
  Since transverse spin is peculiar in having no DIS reference point, the
  effects of higher-order corrections on DY asymmetries are examined via a DIS
  definition for transversity devised using a hypothetical scalar vertex. The
  results suggest that some care may be required when interpreting
  experimentally extracted partonic transversity, particularly when comparing
  with model calculations or predictions.\footnotemark
}%
\footnotetext[1]{%
  Following correction of an error in the code used for the numerical
  estimates, the results shown here are a little less dramatic than those
  actually presented at the symposium.
}%
\section{Motivation}

Transversity is the last, leading-twist piece in the partonic jig-saw puzzle
that makes up the hadronic picture; the theoretical framework (\ie, QCD
evolution, partonic processes, radiative effects, \etc)\ is now rather solid
\cite{Barone:2001sp, Barone:2004p1} while a number of experiments aimed at its
measurement are on-line or under development: HERMES, COMPASS and the RHIC spin
programme. Moreover, transverse-spin effects are notoriously surprising.
\section{Transversity}

\subsection{Chirality and Hikasa's Theorem}

\begin{figure}
  \centering
  \psfrag{(b)}{}
  \hspace*{-3mm}
  \includegraphics[width=40mm,bb=336 571 486 679,clip]{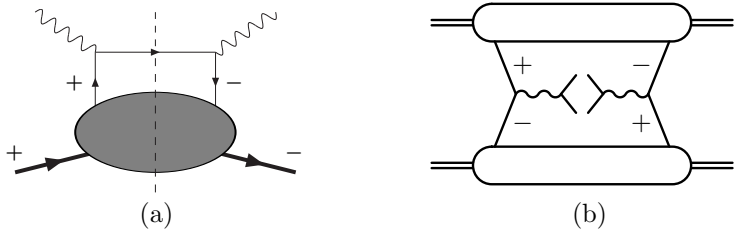}
  \hspace{15mm}
  \mbox{\raisebox{5mm}{%
  \begin{fmfgraph*}(40,25)
    \fmfset{xi}{0.00w}
    \fmfset{xo}{1.00w-xi}
    \fmfset{xc}{0.50w}
    \fmfset{xl}{0.20w}
    \fmfset{xr}{1.00w-xl}
    \fmfset{yc}{0.50h}
    \fmfset{yt}{0.88h}
    \fmfset{yb}{1.00h-yt}
    \fmfset{rx}{0.10h}
    \fmfset{ry}{0.10h}
    \fmfset{ym}{1thin}
    \fmfset{xm}{0.01w}
    \fmfset{xq}{0.07w}
    \fmfset{xp}{xq+0.15w}
    \fmfset{xe}{0.05w}
    \fmfset{ye}{0.10h}
    \fmfi{plain}{(xi,yt+ym)--(xl-rx,yt+ym)}
    \fmfi{plain}{(xi,yt-ym)--(xl-rx,yt-ym)}
    \fmfi{plain}{(xi,yb+ym)--(xl-rx,yb+ym)}
    \fmfi{plain}{(xi,yb-ym)--(xl-rx,yb-ym)}
    \fmfi{plain}{(xo,yt+ym)--(xr+rx,yt+ym)}
    \fmfi{plain}{(xo,yt-ym)--(xr+rx,yt-ym)}
    \fmfi{plain}{(xo,yb+ym)--(xr+rx,yb+ym)}
    \fmfi{plain}{(xo,yb-ym)--(xr+rx,yb-ym)}
    \fmfi{plain}{%
      (xl,yt+ry) -- (xr,yt+ry) .. (xr+rx,yt) ..
      (xr,yt-ry) -- (xl,yt-ry) .. (xl-rx,yt) .. (xl,yt+ry)
    }
    \fmfi{plain}{%
      (xl,yb+ry) -- (xr,yb+ry) .. (xr+rx,yb) ..
      (xr,yb-ry) -- (xl,yb-ry) .. (xl-rx,yb) .. (xl,yb+ry)
    }
    \fmfi{plain,l.d=1mm,         label=$+$}{(xc-xp,yc)--(xl+xm,yt-ry)}
    \fmfi{plain,l.d=1mm,l.s=left,label=$-$}{(xc-xp,yc)--(xl+xm,yb+ry)}
    \fmfi{plain,l.d=1mm,l.s=left,label=$-$}{(xc+xp,yc)--(xr-xm,yt-ry)}
    \fmfi{plain,l.d=1mm,         label=$+$}{(xc+xp,yc)--(xr-xm,yb+ry)}
    \fmfi{photon}{(xc-xp,yc)--(xc-xq,yc)}
    \fmfi{photon}{(xc+xp,yc)--(xc+xq,yc)}
    \fmfi{plain}{(xc-xq,yc)--(xc-xq+xe,yc+ye)}
    \fmfi{plain}{(xc-xq,yc)--(xc-xq+xe,yc-ye)}
    \fmfi{plain}{(xc+xq,yc)--(xc+xq-xe,yc+ye)}
    \fmfi{plain}{(xc+xq,yc)--(xc+xq-xe,yc-ye)}
  \end{fmfgraph*}%
  }}
  \\
  \raisebox{5mm}[0pt][0pt]{(a)\hspace{53mm}(b)\hspace*{0mm}}\\[-2ex]
  \caption{%
    (a) The forbidden chirality-flip DIS handbag diagram.
    (b) The Drell--Yan process for transversely polarised hadrons ($\pm$
    indicate quark chirality).
  }%
  \label{fig:chirality}
\end{figure}

QCD and electroweak vertices conserve quark chirality, so that transversity
decouples from DIS (see Fig.~\ref{fig:chirality}a). Chirality flip is not a
problem if the quarks connect to different hadrons, \eg, as in Drell-Yan (DY)
processes (see Fig~\ref{fig:chirality}b). A caveat to accessing transversity in
DY is \citeauthor{Hikasa:1986qi}'s theorem \cite{Hikasa:1986qi}: chiral
symmetry requires that the lepton-pair azimuthal angle remain unintegrated. No
simple proof exists; it has to do with $\gamma$-matrix properties.

\subsection{Higher-Order Corrections}

Quark densities are usually defined in DIS, where the parton picture was first
formulated and model calculations are performed. When translated to DY, large
radiative $K$ factors appear ${\sim}\Order(\pi\alpha_s)$, enhancing the
cross-section \cite{Altarelli:1979ub}. At RHIC energies the correction is
roughly 30\% while at EMC/SMC energies it becomes nearly 100\%. Since spin
asymmetries are ratios of differences and sums of cross-sections for different
spin-alignment combinations, any strong polarisation dependence in the $K$
factors could lead to dramatic variations in the asymmetries, with respect say
to model predictions. For the $q\bar{q}$ annihilation contribution in the case
of longitudinal polarised hadrons, this turns out \emph{not} to be the case
\cite{Ratcliffe:1983yj}. A partial explanation may be found in the
helicity-conserving nature of vector interactions: only a single helicity
combination contributes, to next-to-leading order (NLO).

However, the case of transversity is peculiar: as noted above, no DIS
definition exists, nor is it obvious that quark helicity-conservation should
still afford any protection. For pure DY, the NLO coefficient functions are
known in various schemes \cite{Vogelsang:1993jn,Contogouris:1994ws};
surprisingly, a new term $\propto\frac{z\ln^2z}{1-z}$ appears, which is found
neither for spin-averaged nor helicity-dependent DY.

Now, to study the $K$-factor problem, we need a DIS-like process to which
transversity may contribute. We thus seek a DIS helicity-flip mechanism, which
could be provided by either a quark mass (\ie, in a propagator) or a scalar
vertex (\eg, a Higgs coupling). Although a quark mass does what is required,
the contribution cancels via the equations of motion and gauge invariance (see,
\eg, \cite{Anselmino:1995gn}). However, a (single) Higgs-like vertex, replacing
one of the photon vertices in Fig.~\ref{fig:chirality}a, allows a chiral-odd
contribution to DIS \cite[from a suggestion by R.L.~Jaffe]{Ioffe:1995aa}.
Indeed, such a \emph{gedanken} process may be used to calculate the anomalous
dimensions, but care is needed.

An attempt at calculating transversity anomalous dimensions $\gamma$ via this
method led to an apparent contradiction, which was corrected by
\citet{Blumlein:2001ca}: the vector current $J_V$ is conserved so $\gamma_V=0$
but the scalar current $J_S$ is not and $\gamma_S\not=0$. The product of two
currents may be expanded as
\begin{equation}
  J_V(\xi) \cdot J_S(0) = \sum_n \, C(n;\xi) \, \Operator(n;0) \,,
\end{equation}
where the RGE's for the Wilson coefficients $C(n;\xi)$ are
\begin{equation}
  \left[
    \CovDer +
    \gamma_{J_V}(g) +
    \gamma_{J_S}(g) -
    \gamma_{\Operator}(n;g)
    \strut
  \right] C(n;\xi)
  = 0 \,.
\end{equation}
Thus, the ``Compton'' amplitude correction has coefficient
\begin{equation}
  \gamma_C(n;g) =
  \gamma_{J_V}(g) +
  \gamma_{J_S}(g) -
  \gamma_{\Operator}(n;g)
\end{equation}
and therefore $\gamma_{\Operator}\not=\gamma_C$\,!

\begin{figure}
  \centering
  \psfrag{DY-2*DIS coefficient difference}
         {\small $C^i_{q,\text{DY}}-2C^i_{q,\text{DIS}}$}
  \psfrag{n}{\small $n$}
  \psfrag{T1}[Br]{\tiny spin averaged\hspace*{-1em}}
  \psfrag{T2}[Br]{\tiny helicity weighted\hspace*{-1em}}
  \psfrag{T3}[Br]{\tiny transversity weighted\hspace*{-1em}}
  \raisebox{-2.1mm}{%
  \includegraphics[width=0.495\textwidth,clip]{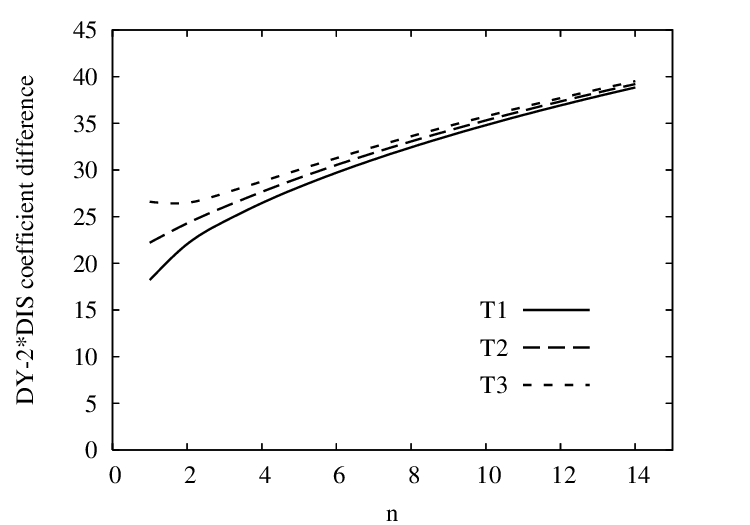}
  }
  \psfrag{A_TT}{\small $A_{TT}$}
  \psfrag{tau}{\small $\tau$}
  \includegraphics[width=0.48\textwidth,clip]{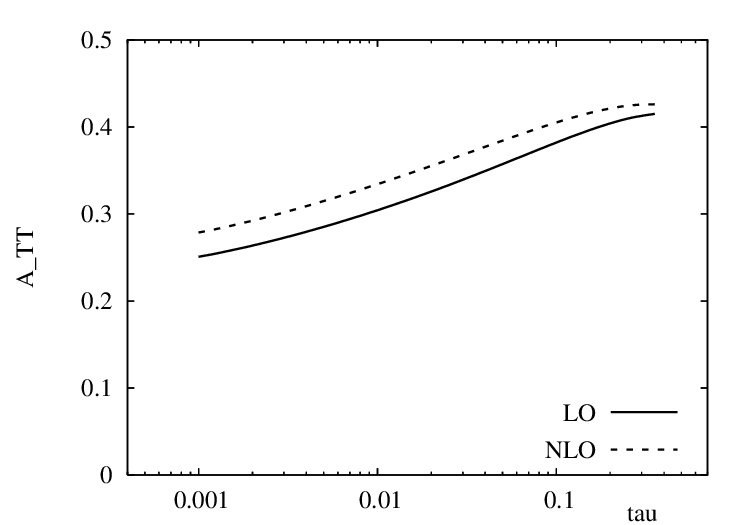}
  \hspace*{-4mm}
  \\
  \raisebox{34mm}[0pt][0pt]{\small (a)\hspace{56mm}(b)\hspace*{24mm}}\\[-2ex]
  \caption{%
    (a)
    Spin-averaged, helicity- and transversity-weighted coefficient
    differences $C^i_{q,\text{DY}}-2C^i_{q,\text{DIS}}$ (for $i=f,g,h$) in
    Mellin moment space.
    (b)
    LO and NLO transversity asymmetries (valence contributions only) for
    Drell--Yan ($\tau=Q^2/s$, $s=1600$\,GeV$^2$).
  }
  \label{fig:curves}
\end{figure}

Moreover, since the scalar current is not conserved, there is an extra UV
contribution from the scalar vertex, which must be factorised into the Higgs
coupling constant (or equivalently, the running quark mass). The results for
the coefficient functions are (see \cite{Ratcliffe:2004a1})
\begin{subequations}
\begin{align}
  C^{f}_{q,\text{DY}} - 2
  C^{f}_{q,\text{DIS}}
  &= \frac{\alpha_s}{2\pi}\CF
  \bigg[
    \frac3{(1-z)_+}
    + 2\left(1+z^2\right) \left(\frac{\ln(1-z)}{1-z}\right)_{\!\!+}
  \nonumber
\\
  & \hspace{7.4em} \null
    - 6 - 4z
    + \left(\frac43\pi^2+1\right)\delta(1-z)
  \bigg] ,
\\[1ex]
  C^{g}_{q,\text{DY}} - 2
  C^{g}_{q,\text{DIS}}
  &=
  C^{f}_{q,\text{DY}} - 2
  C^{f}_{q,\text{DIS}}
  + \frac{\alpha_s}{2\pi}\CF \, 2(1+z) \,,
\\[1ex]
  C^{h}_{q,\text{DY}} - 2
  C^{h}_{q,\text{DIS}}
  &=
  C^{f}_{q,\text{DY}} - 2
  C^{f}_{q,\text{DIS}}
  \nonumber
\\
  & \hspace{3em} \null
  + \frac{\alpha_s}{2\pi}\CF
  \bigg[
    7 - \frac{6z\ln^2z}{1-z} - 2(1-z)\ln(1-z)
  \bigg].
\end{align}
\end{subequations}
The origins of the larger differences in the last line may be traced to
different phase-space restrictions in the transversity case.
Fig.~\ref{fig:curves} shows a comparison of the Mellin moments of the above
coefficients and a simple purely valence estimate of the effects on a
transverse asymmetry. The correction is rather more than twice that of the
helicity case, reaching about 15\%.

It could be argued that it is the Higgs-like vertex that spoils the $K$-factor
cancellation in the transversity case. However, DY processes can also be
constructed in which an intermediate Higgs state produces the lepton pair. The
presence of scalar (chirality-flip) vertices avoids
\citeauthor{Hikasa:1986qi}'s theorem and the final lepton-pair azimuth may be
integrated out. Likewise, a purely Higgs-exchange DIS process exists. In these
cases the large $K$-factors are ``well-behaved''. Thus, model calculations might
not fare too well at first sight if not suitably corrected for the transition
from DIS to DY.
\section{Summary and Conclusions}

A full description of the nucleon must include transversity. On the theory
side, the standard QCD picture is complete to NLO, but only for DY or more
exotic processes. We have no experimental data, though the future is promising.
The phenomenology, while not dissimilar to the other leading-twist densities,
has interesting peculiarities. \citeauthor{Hikasa:1986qi}'s theorem forces us
to keep the lepton-pair azimuth unintegrated in DY, leading to a new term in
the NLO correction, which then affects the $K$-factor. Thus, comparison with
model predictions and even the \citeauthor{Soffer:1995ww} bound
\cite{Soffer:1995ww} could be misleading.

\end{fmffile}
\end{document}